\documentclass[10pt]{iopart}
\usepackage{graphicx}
\usepackage{calc}
\usepackage{SIunits}
\usepackage{textcomp}
\begin{document}
\title{Crystal growth and phase diagram of 112-type iron pnictide superconductor Ca$_{1-y}$La$_{y}$Fe$_{1-x}$Ni$_{x}$As$_{2}$}

\author{Tao Xie$^{1,2}$, Dongliang Gong$^{1,2}$, Wenliang Zhang$^{1,2}$, Yanhong Gu$^{1,2}$, Zita H\"usges$^3$, Dongfeng Chen$^4$, Yuntao Liu$^4$, Lijie Hao$^4$, Siqin Meng$^{4,3}$, Zhilun Lu$^{3}$, Shiliang Li$^{1,2,5}$ and Huiqian Luo$^1$$^{\dag}$}

\address{$^1$ Beijing National Laboratory for Condensed Matter Physics, Institute of Physics, Chinese Academy of Sciences, Beijing 100190, China}
\address{$^2$ University of Chinese Academy of Sciences, Beijing 100190, China}
\address{$^3$ Helmholtz-Zentrum Berlin f\"ur Materialien und Energie GmbH, Berlin 14109, Deutschland}
\address{$^4$ China Institute of Atomic Energy, Beijing 102413, China}
\address{$^5$ Collaborative Innovation Center of Quantum Matter, Beijing 100190, China}

\ead{$^{\dag}$hqluo@aphy.iphy.ac.cn}

\begin{abstract}
We report a systematic crystal growth and characterization of Ca$_{1-y}$La$_{y}$Fe$_{1-x}$Ni$_{x}$As$_{2}$, the newly discovered 112-type iron-based superconductor. After substituting Fe by a small amount of Ni, bulk superconductivity is successfully obtained in high quality single crystals sized up to 6 mm. Resistivity measurements indicate common features for transport properties in this 112-type iron pnictide, suggest strong scattering from chemical dopants. Together with the superconducting transition temperature $T_c$, and the Neel temperature $T_N$ determined by the elastic neutron scattering, we sketch a three-dimensional phase diagram in the combination of both Ni and La dopings.

\end{abstract}

\submitto{\SUST}
\maketitle
\ioptwocol

\section{Introduction}

The superconductivity (SC) discovered in iron pnictides has stimulated intensive researches in condensed matter physics since 2008~\cite{Kamihara} . The basic structure of these materials is a stack of layered square lattices of Fe-As/Fe-P intermediate by alkali / alkali earth / rare-earth ions or oxides. So far, such ingredient has been successfully achieved in many families of iron-based superconductors, which can be classified as: 1111 (e.g. LaFeAsO), 122 (e.g. BaFe$_{2}$As$_{2}$), 111(e.g. LiFeAs),  21311 (e.g. Sr$_2$VO$_3$FeAs), 32522 (e.g. Sr$_3$Sc$_2$O$_5$Fe$_2$As$_2$), 22241(e.g. Ba$_2$Ti$_2$Fe$_2$As$_4$O), 10-3-8 (e.g. Ca$_{10}$(Fe$_3$Pt$_8$)(Fe$_2$As$_2$)$_5$), 10-4-8 (e.g. Ca$_{10}$(Fe$_4$Pt$_8$)(Fe$_2$As$_2$)$_5$), 1144 (e.g. CaRbFe$_4$As$_4$ ), etc ~\cite{Chen X,Chen G,Ren Z,Rotter,Wang X,Hsu,Zhu1,Zhu2,Ogino,Sun Y,Kakiya,Ni1,Iyo1,LiuY,Meier}. Superconductivity emerges from chemical substituted parent compounds with long-ranged antiferromagnetism (AF) or simply in stoichiometric compounds.

\begin{figure}[!ht]\centering
\includegraphics[scale=0.15,clip]{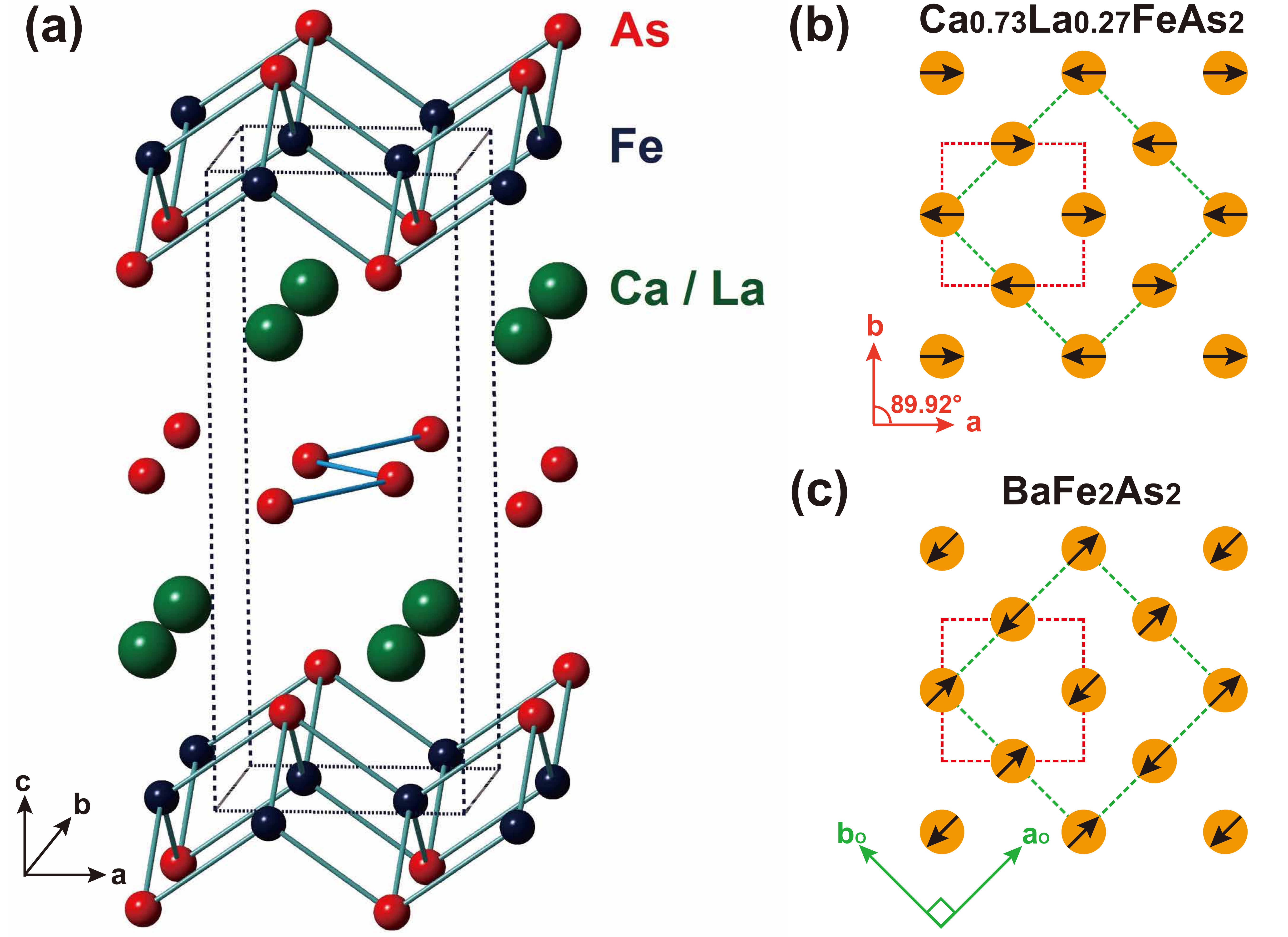}
\caption{(a) Crystal structure of Ca$_{1-y}$La$_{y}$FeAs$_{2}$. (b, c) Comparison of antiferromagnetic structures between Ca$_{0.73}$La$_{0.27}$FeAs$_{2}$ and BaFe$_{2}$As$_{2}$. The red dash box represents the nuclear unit cell, and the green dash box is the magnetic unit cell, respectively. \label{fig:1}}
\end{figure}

In 2013, superconductivity with $T_c$ up to 42 K in a new type iron pnictide compound Ca$_{1-y}$Ln$_{y}$FeAs$_{2}$ (Ln = La, Pr, Nd) was reported and named as 112-type iron-based superconductor~\cite{Katayama,Yakita1}. Although this new system contains similar elements as the tetragonal 1111, 111, and 122 families (space group $I4/mmm$), the crystal structure is monoclinic (space group $P2_{1}$), with a small tilt of the whole quadrangular ($\alpha=90\degree$, $\beta=91.4\degree$, $\gamma=90\degree$) at room temperature and additional zigzag arsenic chains between Ca/La layers~\cite{Katayama,Yakita1,Yakita2} (Fig.\ref{fig:1}(a)). On cooling, the only symmetry of $P2_{1}$ structure will be broken to be triclinic $P1$ by reducing $\gamma$ from $90\degree$ to $89.9\degree$~\cite{Jiang1}, much similar to the 10-3-8 system ~\cite{Ni2}. While the magnetic order, forms just below the structural transition temperature $T_s$, is very similar to co-linear antiferromagnetism in 1111 or 122 type of iron pnictides with wave vector \textbf{Q}$_{AF}$= (1, 0) in orthorhombic lattice~\cite{Dai}, but the ordered moments rotate 45\degree\ away from the stripe directions (Fig.\ref{fig:1}(b)). Such unique features result in a structurally untwinned lattice but twinned magnetic domains at low temperature for the weak magneto-elastic coupling~\cite{Jiang1}. Unfortunately, just like the rare-earth doped CaFe$_2$As$_2$ ~\cite{Lv}, only filamentary superconductivity can be found in Ca$_{1-y}$Ln$_{y}$FeAs$_{2}$ compound. Though the iso-valent substitution of As by Sb/P or  post-annealing of the crystals may improve the superconducting transition and volume, it is still hard to obtain bulk superconductivity with full Meissner shielding volume~\cite{Katayama,Yakita1,Yakita2,Kudo}. Recently, Jiang \emph{et. al. } have finally realized bulk superconductivity by slightly doping Co into the \textquotedblleft parent\textquotedblright compound Ca$_{0.73}$La$_{0.27}$FeAs$_{2}$ with $T_s=$ 58 K and $T_N=$ 54 K~\cite{Jiang2}. Both antiferromagnetic and superconducting phases coexist microscopically with each other in the underdoped region, and optimal superconductivity with maximum $T_c= 20 $ K is obtained at 4.6\% Co doping level. Surprisingly, nuclear magnetic resonance (NMR) research reveals that lanthanum doping, which is generally believed to introduce electrons~\cite{Chen X}, will actually enhance $T_N$ but suppress $T_c$ simultaneously~\cite{Kawasaki}. To understand the magnetism and superconductivity in this fascinating material, it is necessary to further establish the detailed phase diagram concerning different chemical dopings both at Ca and Fe sites.

In this article, we report our results of crystal growth and characterization of Ca$_{1-y}$La$_{y}$Fe$_{1-x}$Ni$_{x}$\-As$_{2}$ with various Ni doping levels ($0 \leq x \leq 0.24$) and La doping ($y=$ 0.18 and 0.24). Most of them show bulk supercondutivity at the superconducting state, and systematic evolution of transport properties at the normal state. With the $T_N$ determined from neutron diffraction experiments on the single crystals and previous report~\cite{Jiang1}, and $T_c$ obtained from magnetization and resistivity measurements, we establish a three-dimensional phase diagram of Ca$_{1-y}$La$_{y}$Fe$_{1-x}$Ni$_{x}$\-As$_{2}$ with rich interplay between $T_c$ and $T_N$.

\section{Experiment}

\begin{figure}[!ht]\centering
\includegraphics[scale=0.3,clip]{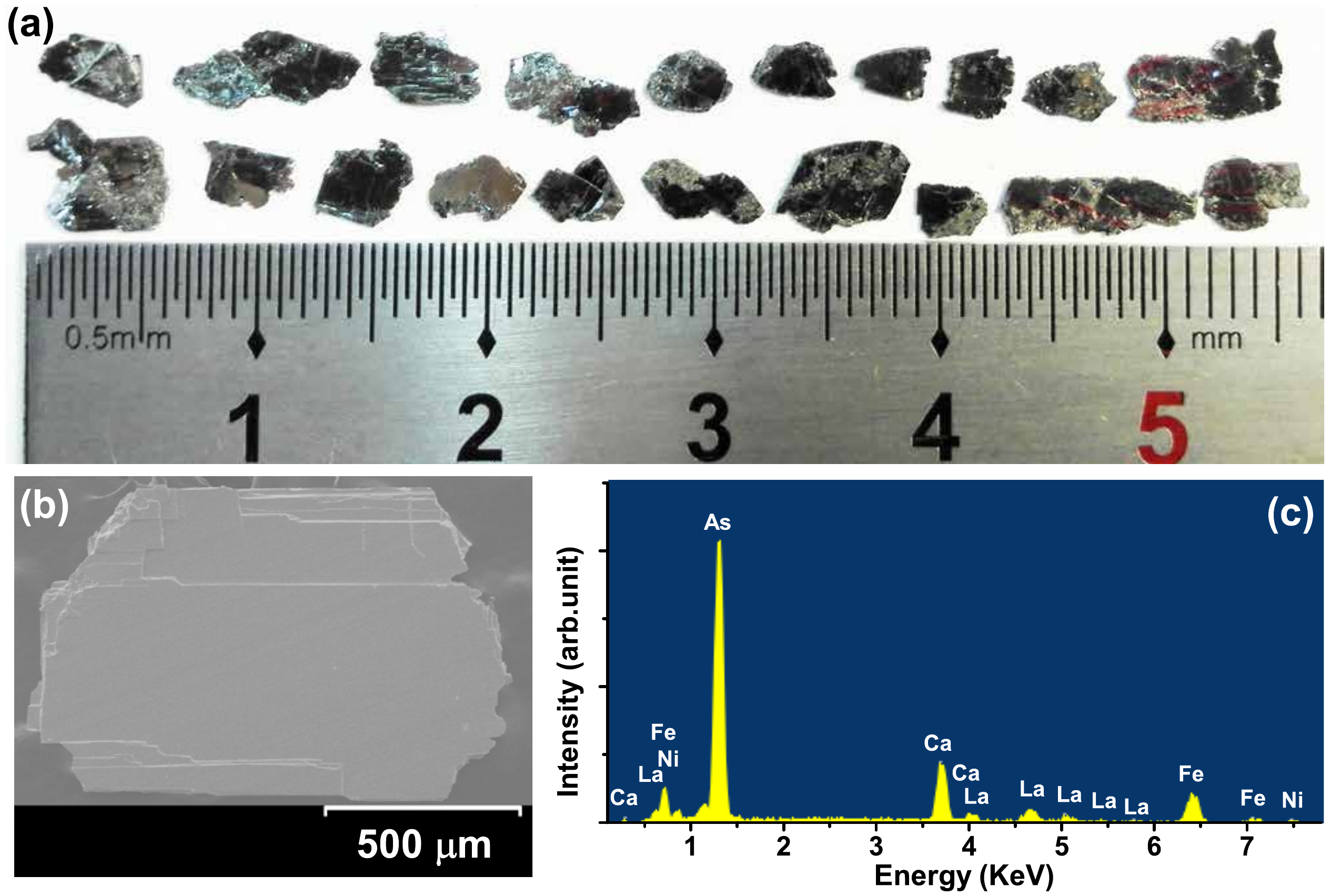}
\caption{(a) Photo of the Ca$_{0.82}$La$_{0.18}$Fe$_{1-x}$Ni$_{x}$As$_{2}$ crystals. (b, c) SEM image and EDX spectrum of one typical crystal.\label{fig:2}}
\end{figure}

 Ca$_{1-y}$La$_{y}$Fe$_{1-x}$\-Ni$_{x}$\-As$_{2}$ single crystals were grown by self-flux method using CaAs as flux~\cite{Luo H}. Before the crystal growth, the precursors CaAs, LaAs, Fe$_{1-x'}$Ni$_{x'}$As (in nominal composition) had been prepared with raw materials Ca(Alfa Aesar, $>$99.5\%), La(Alfa Aesar, $>$99.9\%), Fe(Alfa Aesar, $>$99.5\%), Ni(Alfa Aesar, $>$99.99\%), As(Alfa Aesar, $>$99.99\%) by solid state reaction method. For CaAs, the Ca granular and ground As chips with a ratio of 1 : 1 were sealed into an evacuated quartz tube and placed in a box furnace. The raw materials were heated to 400\degreecelsius{} slowly and held for more than 10 hrs, then further heated to 630\degreecelsius{} slowly and held for another 20 hrs, and finally quenched to room temperature. The mixture for the first batch was ground into powder, sealed into an evacuated quartz tube, and heated up to 670\degreecelsius{} again for complete reaction. For LaAs, the starting materials La and As chips were mixed with a ratio of 1 : 1 and loaded into an Al$_{2}$O$_{3}$ crucible and sealed into a quartz tube. The whole ampoule was slowly heated to 500\degreecelsius{} in 20 hrs, held for 10 hrs, then heated to 850\degreecelsius{} in 10 hrs and held for another 20 hrs. The Fe$_{1-x'}$Ni$_{x'}$As powders were prepared by the same method we used before~\cite{Chen Y}. All precursors were checked to be high pure crystalline phases by X-ray powder diffraction. Finally, for the Ca$_{0.82}$La$_{0.18}$\-Fe$_{1-x}$Ni$_{x}$As$_{2}$ samples, precursors CaAs, LaAs, Fe$_{1-x'}$Ni$_{x'}$As with a molar ratio of 3.7 : 0.3 : 1 and total mass about 10 grams were grounded to mix up homogeneously, then pressed into pellets, loaded into a $\phi 21$ mm $\times$ 60 mm Al$_{2}$O$_{3}$ crucible and sealed into a quartz tube with inner size $\phi 23$ mm $\times$ 100 mm. For the Ca$_{0.76}$La$_{0.24}$\-Fe$_{1-x}$Ni$_{x}$As$_{2}$ samples, we grow the crystals by changing the molar ratio to 1.65 : 0.35 : 1 for the corresponding precursors CaAs, LaAs, Fe$_{1-x'}$Ni$_{x'}$As. The ampoule was heated to 900\degreecelsius{} in 15 hrs and kept for 10 hrs first, then heated to 1180\degreecelsius{} in 5 hrs and held for 20 hrs for melting. At the last stage, the furnace temperature was decreased to 950\degreecelsius{} at a slow rate of 3\degreecelsius{}  / h and then quenched to room temperature. By cracking the melted pellet after exposing them in the air for several hours, sizable shining plate-like single crystals were successfully obtained.

\begin{figure}[!ht]\centering
\includegraphics[scale=0.16,clip]{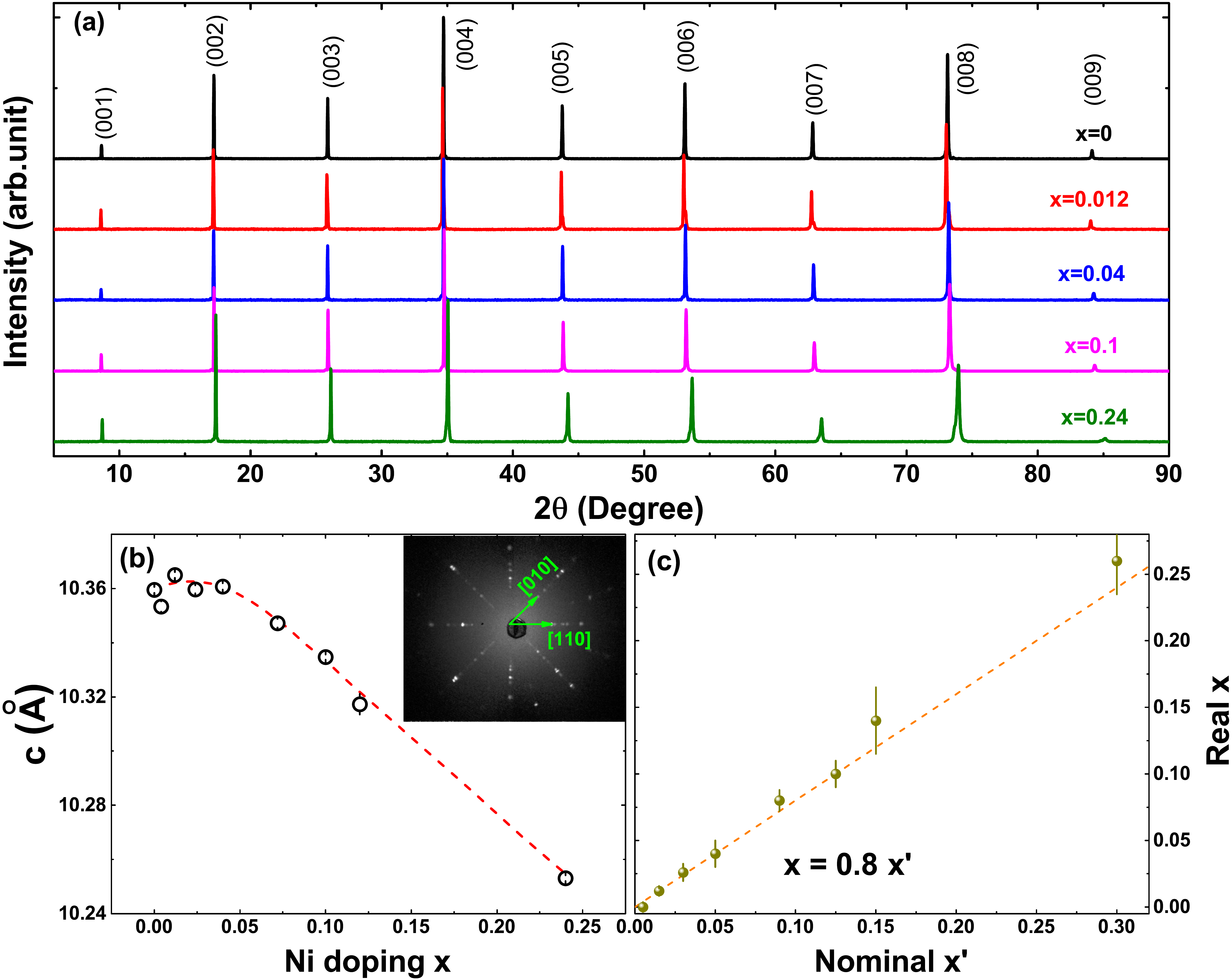}
\caption{(a) X-ray diffraction patterns of Ca$_{0.82}$La$_{0.18}$\-Fe$_{1-x}$Ni$_{x}$As$_{2}$ single crystals at room temperature. (b) The Ni doping evolution of $c$-axis, where the insert shows the Laue photo of one typical crystal. (c) Linear relationship between the nominal Ni concentration $x'$ and the real Ni concentration $x$. The dash line is a straight line with $x$ = 0.8 $x'$.\label{fig:3}}
\end{figure}

To check the quality of our crystals, the crystallinity, chemical composition and planeness of cleaved surface were characterized. Single crystals X-ray diffraction (XRD) were carried out on a SmartLab 9 kW high resolution diffraction system with Cu K$\alpha$ radiation($\lambda$ = 1.540598 \angstrom) at room temperature ranged from $5\degree$ to $90\degree$ in refection mode. The Laue photos of the crystals were taken by a Photonic Sciences Laue camera in backscattering mode with incident beam along $c$-axis. The microscopic morphology and energy-dispersive X-ray spectrum (EDX) of the crystals were measured by a high resolution cold field emission scanning electron microscope (SEM) (Hitachi S-4800) equipped with an energy dispersive X-ray spectrometer. The accurate composition of our samples were determined by the inductively coupled plasma (ICP) analysis.

The superconducting transition temperatures were determined by DC magnetization measurements on a \emph{Quantum Design} Magnetic Property Measurement System (MPMS) with zero-field-cooling method and $H \parallel ab$ plane, where the demagnetization factor is nearly zero for the very thin crystals.  The superconducting and normal state transport properties were further measured by standard four-probe resistance measurements on a \emph{Quantum Design} Physical Property Measurement System (PPMS) down to 2 K. Four Ohmic contacts were painted by DuPont 5025 silver paint on the crystals with contact resistance less than 1 $\Omega$. Temperature dependence of resistivity was measured by sweeping temperature in a low rate of 1 K/min. For each doping, we measured at least 3 typical pieces of crystals to repeat the results.

The antiferromagnetism on several samples were measured by elastic neutron scattering experiments on the cold neutron triple-axis spectrometer FLEXX-V2 at BER II, Helmholtz-Zentrum Berlin (HZB) in Germany. The fixed final
energy was $E_f =$ 5 meV (in wavelength $\lambda_f=$ 4.05 \angstrom) with a velocity selector before the sample and a cold Be filter after the sample. The wave vector {\bf Q} at ($q_x$, $q_y$, $q_z$) was defined as ($H$,$K$,$L$) = ($q_x$a/2$\pi$, $q_y$b/2$\pi$, $q_z$c/2$\pi$) reciprocal lattice units(r.l.u.) by simply using the tetragonal notation  where a $\approx$ b $\approx$ 3.90 \angstrom{ }and c $\approx$ 10.31 \angstrom. All the samples were aligned in [$H$, $H$, 0] $\times$ [0, 0, $L$] scattering plane, to reach the magnetic wave vector \textbf{Q}$_{AF}$= (0.5, 0.5, $L/2$) with $L=\pm 1, \pm 2, \pm 3,...$ .

\section{Result and discussion}

\begin{figure}[!ht]\centering
\includegraphics[scale=0.3,clip]{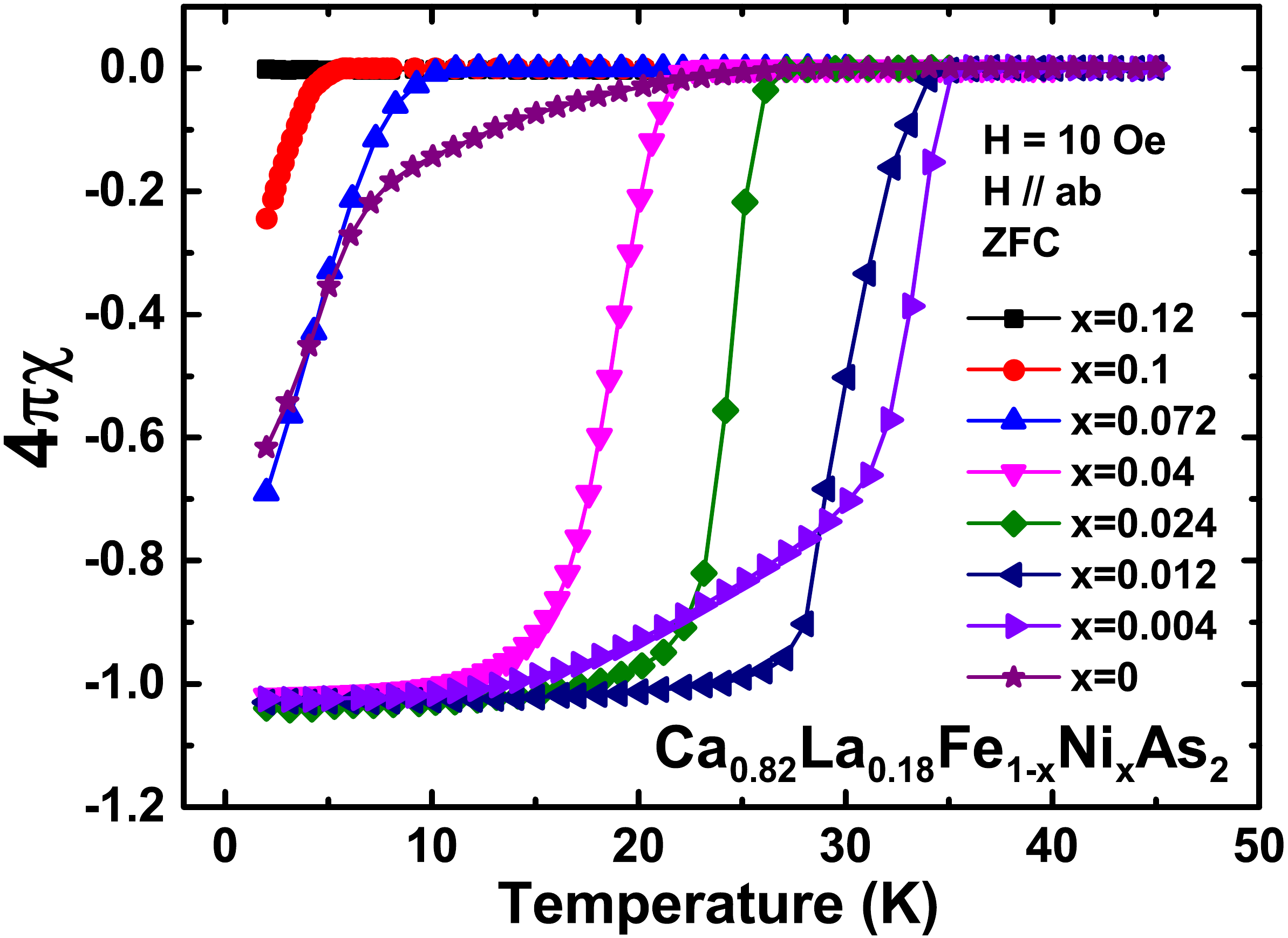}
\caption{The DC magnetic susceptibility for Ca$_{0.82}$La$_{0.18}$Fe$_{1-x}$Ni$_{x}$As$_{2}$ single crystals.\label{fig:4}}
\end{figure}

\begin{figure}[!ht]\centering
\includegraphics[scale=0.25,clip]{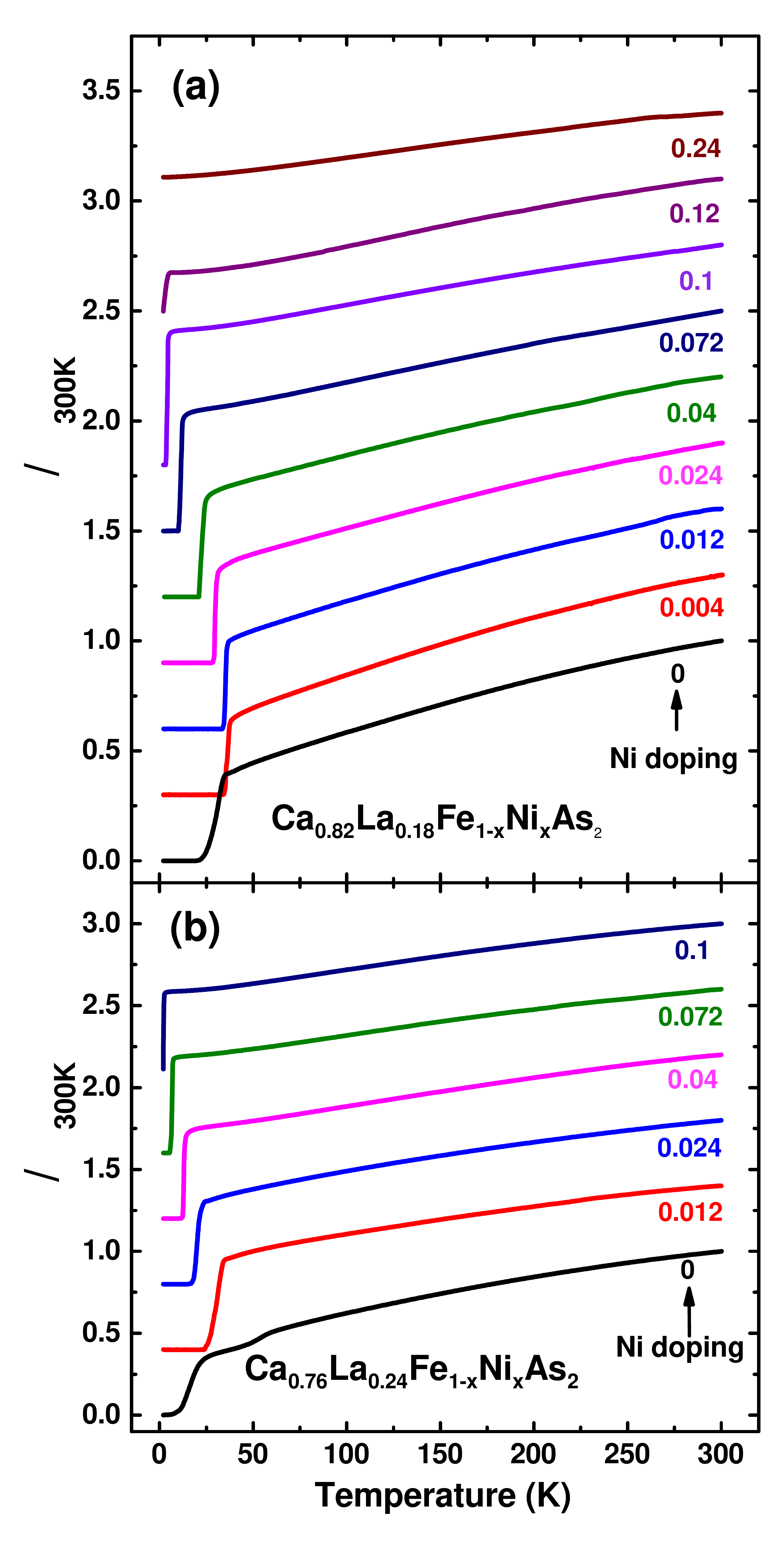}
\caption{Temperature dependence of the resistivity for Ca$_{1-y}$La$_{y}$Fe$_{1-x}$\-Ni$_{x}$\-As$_{2}$ single crystals. All data is normalized by the resistivity at 300 K and shifted upward with 0.3 and 0.4 one by one for (a) and (b), respectively. \label{fig:5}}
\end{figure}

\begin{figure}[!ht]\centering
\includegraphics[scale=0.3,clip]{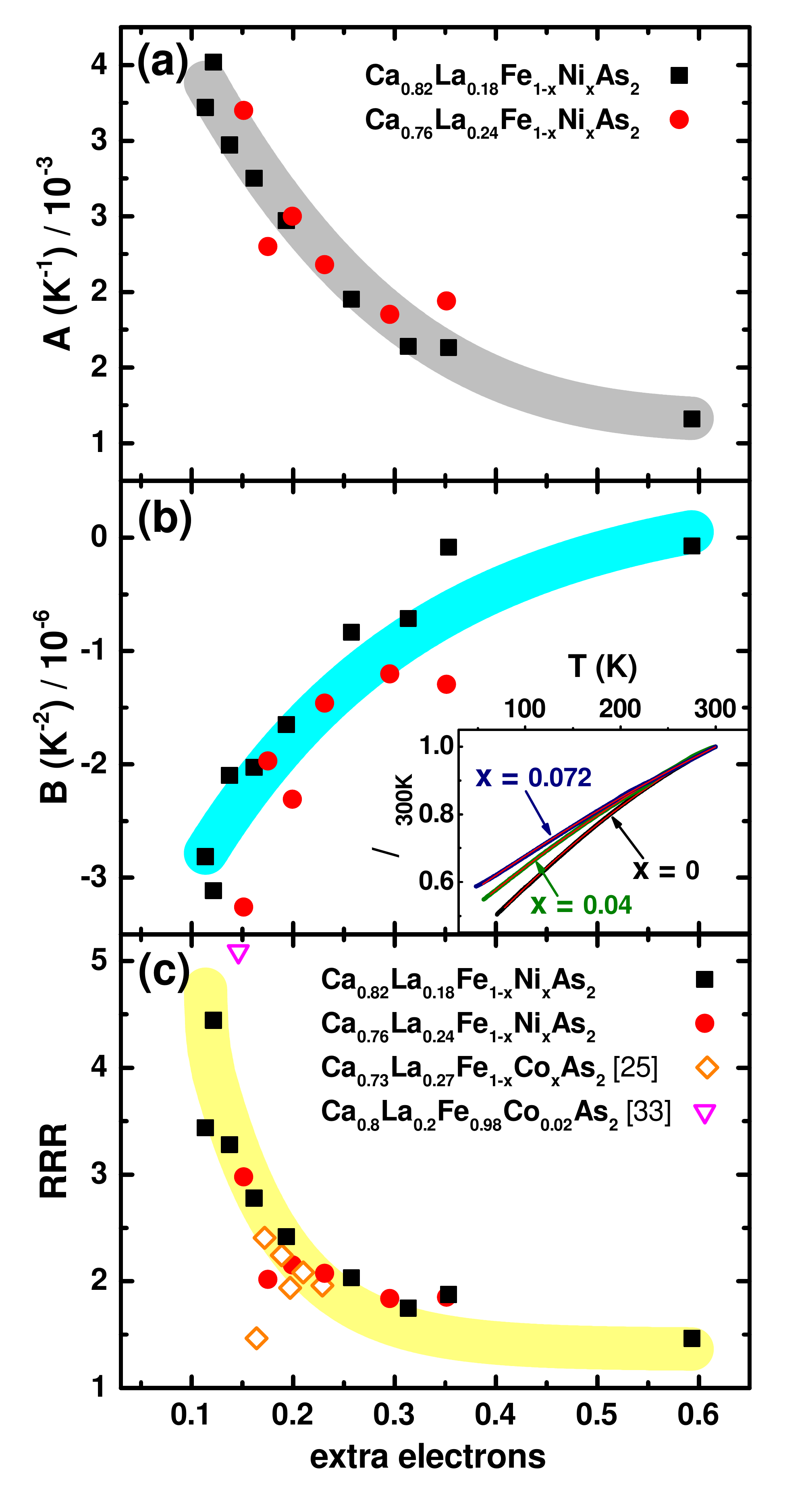}
\caption{(a, b) The extra electrons dependence of the coefficient A, B from the fitting $\rho(T)/\rho$(300 K) = {$\it\rho$}$_{0}$ + A{\it T} + B{\it T}$^2$ of Ca$_{1-y}$La$_{y}$\-Fe$_{1-x}$Ni$_{x}$As$_{2}$. Insert of (b) shows the typical fitting results for Ca$_{0.82}$La$_{0.18}$\-Fe$_{1-x}$Ni$_{x}$As$_{2}$ with $x=$ 0, 0.04, and 0.072 at normal state, where the red lines are fitting curves. (c) The extra electrons dependence of RRR of Ca$_{1-y}$La$_{y}$Fe$_{1-x}$\-{\it TM}$_{x}$\-As$_{2}$ ({\it TM} = Ni, Co). The extra electrons are defined as $2x + 0.63y$ for Ca$_{1-y}$La$_{y}$\-Fe$_{1-x}$Ni$_{x}$As$_{2}$ and $x + 0.63y$ for Ca$_{1-y}$La$_{y}$\-Fe$_{1-x}$Co$_{x}$As$_{2}$, respectively. The shadow areas are guides for the eyes.\label{fig:6}}
\end{figure}

We have successfully grown two groups of Ni doped 112-type iron pnictides with different La contents: Ca$_{0.82}$La$_{0.18}$\-Fe$_{1-x}$Ni$_{x}$As$_{2}$ and Ca$_{0.76}$La$_{0.24}$\-Fe$_{1-x}$Ni$_{x}$As$_{2}$. The crystallographic results are quite similar for the two series, thus we only show typical results of the former one with La concentration $y=$ 0.18. Figure \ref{fig:2}(a) shows the as-grown crystals of Ca$_{0.82}$La$_{0.18}$\-Fe$_{1-x}$Ni$_{x}$As$_{2}$ with the largest size 5 \texttimes{} 6 \texttimes{} 0.5 mm$^3$. All crystals have shiny surface after crashing from the ingot in the crucible. The SEM image in Fig.\ref{fig:2}(b) shows the detailed characteristic of the crystal with some naturally cleaved edge along [1, 1, 0] direction, as determined by X-ray Laue reflection (Fig.\ref{fig:3}(c)). Figure \ref{fig:2}(c) gives the EDX spectrum of the same crystal in Fig.\ref{fig:2}(b), all elements including Ca, La, Fe, Ni, As can be detected, and their contents can be roughly estimated from the spectrum weight. To check the crystalline quality, we have performed single crystal XRD measurements on each doping level at room temperature. Five typical XRD patterns are presented in Fig.\ref{fig:3}(a) for Ca$_{0.82}$La$_{0.18}$\-Fe$_{1-x}$Ni$_{x}$As$_{2}$ with $x=$ 0, 0.012, 0.04, 0.1, 0.24. The sharp (0 0 {\it l}) peaks indicate high $c$-axis orientation of our crystals. No 122 phase  of Ca$_{1-y}$La$_{y}$\-Fe$_{2-x}$Ni$_{x}$As$_{2}$, which has larger lattice parameter $c$ and can be only indexed by even peaks along $c$-axis, has been found in all examined samples. The slightly shift of Bragg peaks toward to high 2$\theta$ angles indicates the decreasing of the length of $c$-axis with Ni doping level increasing, as summarized in Fig.\ref{fig:3}(b). The chemical composition of our samples are characterized by ICP analysis with about 1\% uncertainty. If we suppose the nominal Ni doping level $x'$ in precursor Fe$_{1-x'}$Ni$_{x'}$As, then we have a linear relation between the real content of Ni $x$ and the nominal one $x'$, $x$ = 0.8 $x'$, where the error bars in Fig.\ref{fig:3}(c) come from the statistics among 3-5 pieces of crystal in the same batch. Such segregation coefficient is same as the case of Ni doped BaFe$_2$As$_2$ ~\cite{Luo H,Chen Y,Zhang1,Ni3}, suggesting common ability of congruent melting between Fe and Ni in iron pnictides. The systematic evolution of the $c$-axis and real compositions prove the reliable and repeatable of our grown method. In order to compare with the previous results, we will use the real composition $x$ in the following discussion.

\begin{figure*}[!ht]\centering
\includegraphics[scale=0.45,clip]{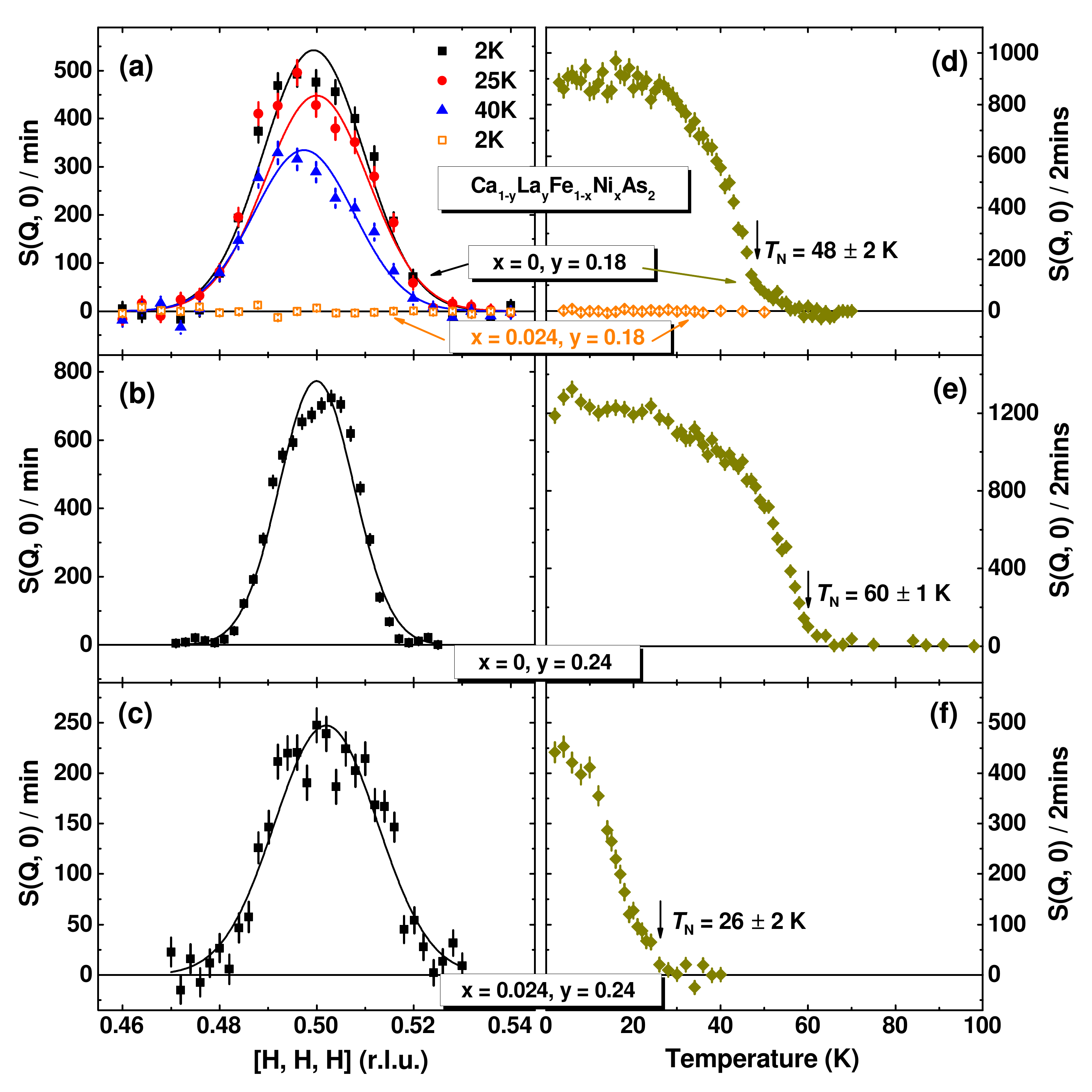}
\caption{Neutron diffraction experiments on Ca$_{1-y}$La$_{y}$Fe$_{1-x}$\-Ni$_{x}$\-As$_{2}$ single crystals. (a - c) Magnetic Bragg peaks by scaning $Q$ along [$H$, $H$, $H$] direction for $x =$ 0, $y =$ 0.18 and $x =$ 0.024, $y =$ 0.18; $x =$ 0, $y =$ 0.24; $x =$ 0.024, $y =$ 0.24. The solid lines in the figures are gauss fittings of the data. (d - f) Temperature dependence of the magnetic scattering at {\bf Q}$_{AF}$ = (0.5, 0.5, 0.5) for the corresponding doping in (a - c). The antiferromagnetic phase transition temperature $T_N$ is marked by the black arrows.\label{fig:7}}
\end{figure*}

\begin{figure*}[!ht]\centering
\includegraphics[scale=0.34,clip]{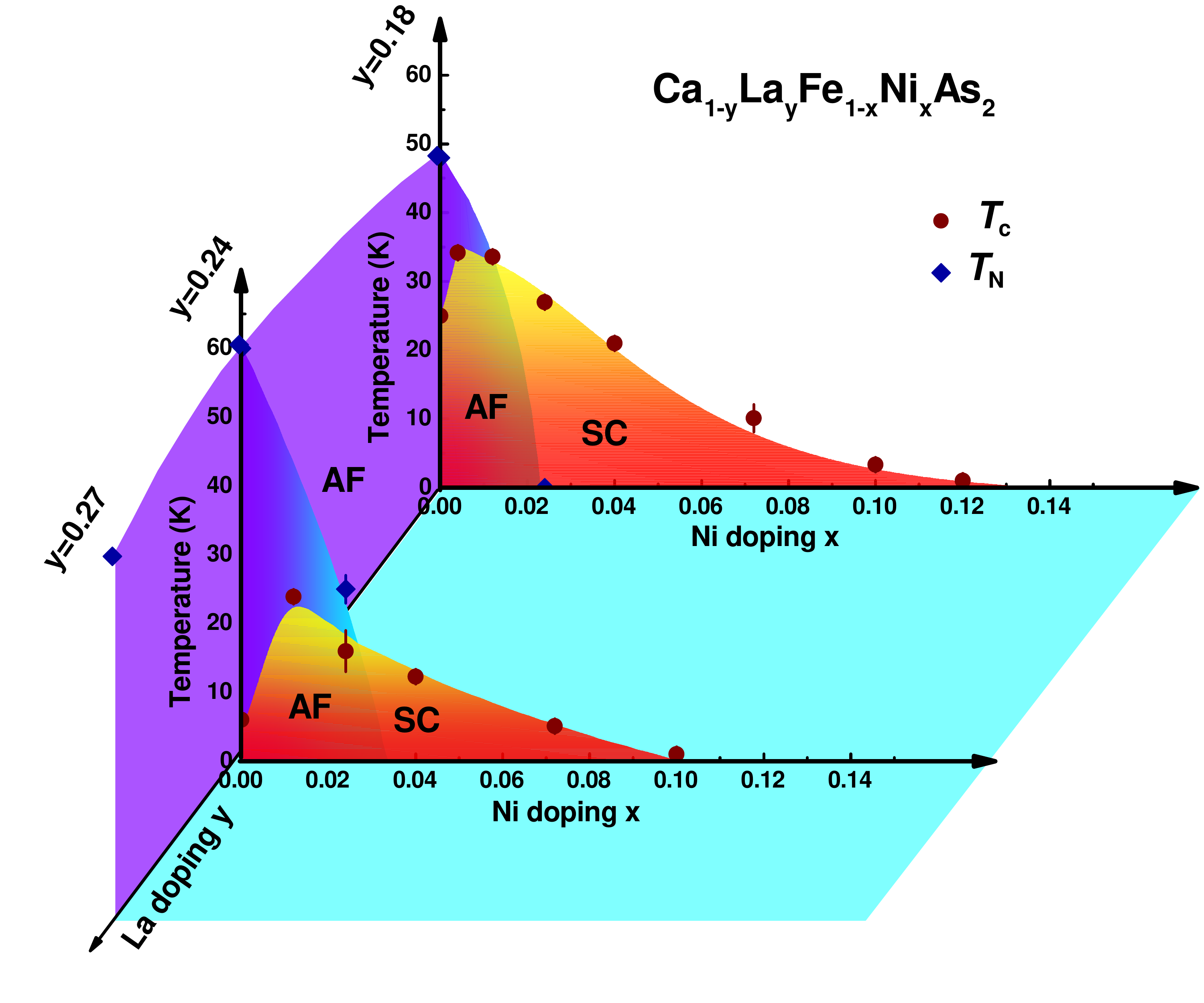}
\caption{The 3D phase diagram of Ca$_{1-y}$La$_{y}$Fe$_{1-x}$\-Ni$_{x}$\-As$_{2}$, where the AF, SC represent antiferrmagnetism and superconductivity with transition temperature $T_N$ and $T_c$, respectively. \label{fig:8}}
\end{figure*}

Figure \ref{fig:4} shows the temperature dependence of the DC magnetic susceptibility for  Ca$_{0.82}$La$_{0.18}$\-Fe$_{1-x}$Ni$_{x}$As$_{2}$ with $x=0 \sim 0.12$. For the Ni-free Ca$_{0.82}$La$_{0.18}$\-FeAs$_{2}$, the onset diamagnetism appears at 25 K with a very broad superconducting transition, and the diamagnetism does not saturate at 2 K. With very little Ni doping, the superconducting transition temperature $T_c$ can be significantly improved from 25 K for $x= 0$ to 34 K for $x=$ 0.004, then slowly drop down upon further Ni doping. This may indicate that this system is very close to an optimal condition for superconductivity. More impressively, after dopping more Ni, the superconducting transition width become very sharp and most of the superconducting samples have nearly full Meissner shielding volume ($4\pi\chi\approx -1$) at base temperature($T = 2$ K). Upon further doping, this system turns to be non-superconducting over $x=$ 0.12 for heavily overdoped electrons.

The normalized temperature dependence of resistivity up to 300 K $\rho(T)/\rho$(300 K) are presented in Fig.\ref{fig:5} both for Ca$_{0.82}$La$_{0.18}$Fe$_{1-x}$Ni$_{x}$As$_{2}$ and Ca$_{0.76}$La$_{0.24}$Fe$_{1-x}$Ni$_{x}$As$_{2}$. The systematic evolution of superconductivity is very clear for both groups of our samples. The normal state of all dopings behaviors like a metal similar to the BaFe$_{2-x}$Ni$_{x}$As$_{2}$ system ~\cite{Chen Y,Ni3,Zhang2}. To describe the electronic transport properties, we fit the $\rho(T)/\rho$(300 K) data by an empirical formula in the combination of linear and quadratic components: $\rho(T)/\rho$(300 K) $= \rho_0 + AT + BT^2$ within a wide temperature range from 30 K above the $T_N$ or $T_c$ to 300 K, which is commonly used in some cuprates, pnictides and organic superconductors~\cite{Analytis}. Generally, the coefficient A represents the non-Fermi-liquid behaviors, while the coefficient B means the proportion of Fermi-liquid behaviors, and $\rho_0$ is the normalized residual resistivity. In this way, we can also obtain the residual resistivity ratio RRR = $\rho$(300 K)/$\rho$(0 K) = 1/$\rho_0$ related to the strength of impurity scattering. For both cases with $y=$ 0.18 and 0.24, such fitting agrees very well up to 300 K. Similarly, we can calculate the magnitude of RRR in Ca$_{0.74}$La$_{0.26}$Fe$_{1-x}$Co$_{x}$As$_{2}$ and Ca$_{0.8}$La$_{0.2}$Fe$_{0.98}$Co$_{0.02}$As$_{2}$ compounds reported before~\cite{Jiang2,Xing X}. By simply considering both La and Ni/Co dopants contribute electrons into the system, e.g., the DMFT calculation indicates 0.17 e/Fe in the Fe$_{2}$As$_{2}$ layer for Ca$_{0.73}$La$_{0.27}$FeAs$_{2}$ compound~\cite{Jiang1}, we could unify the electron doping level to be extra electrons, with 0.63 e/Fe per La and 2 e/Fe per Ni or 1 e/Fe per Co. Here, the value of extra electrons equal to $2x + 0.63y$ for Ca$_{1-y}$La$_{y}$\-Fe$_{1-x}$Ni$_{x}$As$_{2}$ and $x + 0.63y$ for Ca$_{1-y}$La$_{y}$\-Fe$_{1-x}$Co$_{x}$As$_{2}$, respectively. We thus summary the fitting parameters $A$, $B$, and RRR versus extra electrons in Fig. \ref{fig:6}. Interestingly, $A$ and $B$ show opposite behaviors upon electron doping, suggesting the transport behaviors become more like Fermi liquid with the increasing extra electrons, much similar to the overdoped case in BaFe$_{2-x}$Ni$_{x}$As$_{2}$~\cite{Zhang2}, which is consistent with previous reports\cite{Jiang1,Jiang2}. The rapidly decreasing of A also supports that the Ni-free compounds are nearly optimized for superconductivity. It should be noticed that RRR scales very well with extra electrons, no matter the specific chemical dopants. Indeed, higher electron doping level means stronger scattering from the impurity effect and larger $\rho_0$. This probably is the cause of the quick $T_c$ reduction for $x\geq 0.012$ in Ca$_{0.82}$La$_{0.18}$Fe$_{1-x}$Ni$_{x}$As$_{2}$.

By a careful inspection of the data in Fig.\ref{fig:5}, one can find resistivity anomalies in Ni-free samples Ca$_{0.82}$La$_{0.18}$FeAs$_{2}$ and Ca$_{0.76}$La$_{0.24}$FeAs$_{2}$ around 50 K and 60 K, respectively. The anomalies disappear immediately after substituting Ni. Such features may be induced either by structural transition or magnetic phase transition, similar to the case in Ca$_{0.73}$La$_{0.27}$FeAs$_{2}$~\cite{Jiang1}. Therefore, we perform elastic neutron scattering experiments on several compounds of Ca$_{1-y}$La$_{y}$Fe$_{1-x}$Ni$_{x}$As$_{2}$ with $x$ = 0, $y$ = 0.18; $x$ = 0.024, $y$ = 0.18; $x$ = 0, $y$ = 0.24; $x$ = 0.024, $y$ = 0.24.  Temperature dependence of nuclear scattering at {\bf Q} = (1, 1, 0) is done to search possible structural transition, NO clear peak splitting OR extinction effects are found for all samples within the spectrometer resolution, indicating very weak lattice distortion in these compounds. Then magnetic scattering in {\bf Q}-scans along [$H$, $H$, $H$] direction ($\theta-2\theta$ scans) at different temperatures are measured, as shown in Fig.\ref{fig:7}(a-c). Resolution limited peaks are mapped out centering around the wave vector {\bf Q}$_{AF}$=(0.5, 0.5, 0.5) for the samples with $x$ = 0, $y$ = 0.18; $x$ = 0, $y$ = 0.24; $x$ = 0.024, $y$ = 0.24, suggesting long-ranged stripe magnetic order in them. There is no detectable magnetic Bragg peak in the compound with $x$ = 0.024, $y$ = 0.18 (Fig.\ref{fig:7}(a,d)). The order parameter of the antiferromagnetism can be characterized by the temperature dependence of the intensity of magnetic Bragg peak, as correspondingly shown in Fig.\ref{fig:7}(d-f). The Neel temperature $T_N$ are determined to be 48 $\pm$ 2 K , 60 $\pm$ 1 K, 26 $\pm$ 2 K, and the ordered moments are 0.27 $\pm$ 0.1 $\mu$$_{B}$ , 0.33 $\pm$ 0.05 $\mu$$_{B}$ , 0.09 $\pm$ 0.01 $\mu$$_{B}$ for the three samples $x$ = 0, $y$ = 0.18; $x$ = 0, $y$ = 0.24; $x$ = 0.024, $y$ = 0.24 , respectively. The $T_N$ determined here for Ca$_{0.82}$La$_{0.18}$FeAs$_{2}$ and Ca$_{0.76}$La$_{0.24}$FeAs$_{2}$ are consistent with the temperatures of the resistivity anomalies. Therefore, the Ni dopings indeed suppress $T_N$ rapidly in Ca$_{1-y}$La$_{y}$Fe$_{1-x}$Ni$_{x}$As$_{2}$ system, while the La doping have weak effect on the magnetic order.

Finally, together with all results from resistivity, magnetic susceptibility and neutron diffraction experiments, as well as previous report~\cite{Jiang1} on Ca$_{0.73}$La$_{0.27}$FeAs$_{2}$ compound with $T_N=$ 54 K, we sketch a three-dimensional phase diagram for Ca$_{1-y}$La$_{y}$Fe$_{1-x}$Ni$_{x}$As$_{2}$ system,  as shown in Fig.\ref{fig:8}. Some conclusions can be found in this phase diagram. First, by fixing the La doping, the electron doping from Ni quickly suppress the magnetic order and introduce a superconducting dome with co-existing region with AF. Second, in the non-Ni-doped Ca$_{1-y}$La$_{y}$FeAs$_{2}$, La doping can slightly change the $T_N$ around 55 $\pm$ 5 K. Third, the higher La doping level give smaller superconducting regime with further Ni doping because of higher defect/impurity scattering in the samples, but extend the co-existing region of AF and SC. Therefore, the effects from La and Ni dopings are more complicate than those cases simply introducing electrons to suppress the AF and induce SC in other iron pnictides.

\section{Summary}

In summary, high-quality sizable single crystals of Ca$_{1-y}$La$_{y}$Fe$_{1-x}$Ni$_{x}$As$_{2}$ have been grown successfully by the self-flux method. Transport behaviors at normal state can be simply described by doped extra electrons in the 112-type families. The detailed phase diagram shows that bulk superconductivity can be introduced by Ni doping above $x=0.004$ after suppressing the long-ranged AF order. Although the La doping have weak effect on the magnetism of Ni-free Ca$_{1-y}$La$_{y}$FeAs$_{2}$, larger co-existing regime between SC and AF but narrower superconducting dome are obtained for $y=0.24$ system rather than $y=0.18$ case due to stronger impurity scattering. Such rich phenomena on the electronic states opens a new opportunity to reveal the origin of the magnetism and superconductivity in iron pnictides.

\section*{Acknowledgments}
This work is supported by the National Natural Science
Foundation of China (Nos. 11374011, 11374346, 11674406 and 11674372), the Strategic Priority Research
Program (B) of the Chinese Academy of Sciences (XDB07020300), the Ministry of Science and
Technology of China (No. 2016YFA0300502), and the Youth Innovation Promotion Association of CAS (No. 2016004).

\end{document}